\documentclass[12pt,a4paper]{article}    

\usepackage[margin=1in]{geometry} 
\usepackage{amsmath,amssymb,amsthm}
\usepackage{color}
\usepackage{relsize}
\usepackage[perpage,symbol*]{footmisc}
\usepackage{authblk}
\usepackage{enumerate}
\usepackage{cases}
\usepackage{cite}

\newtheorem{theorem}{Theorem}[section]
\newtheorem{proposition}[theorem]{Proposition}
\newtheorem{lemma}[theorem]{Lemma}

\newtheorem{remark}{Remark}

\newcommand{\bt}{\beta}

\newcommand{\be}{\begin{equation}}
\newcommand{\ee}{\end{equation}}
\newcommand{\bea}{\begin{eqnarray}}
\newcommand{\eea}{\end{eqnarray}}
\newcommand{\e}{{\rm e}}

\numberwithin{equation}{section}

\title{Hankel Determinants for a Gaussian weight with Fisher-Hartwig Singularities and Generalized\\ Painlev\'{e} IV Equation}
\author[1]{Xinyu Mu}
\author[1,\thanks{Author to whom any correspondence should be addressed. Email: lvshulin1989@163.com}]{Shulin Lyu}
\affil[1]{School of Mathematics and Statistics, Qilu University of Technology (Shandong Academy of Sciences), Jinan
250353, China}

\date{\today}

\begin{document}
\maketitle

\begin{abstract}	
We study the Hankel determinant generated by a Gaussian weight with Fisher-Hartwig singularities of root type at $t_j$, $j=1,\cdots ,N$. It characterizes a type of average characteristic polynomial of matrices from Gaussian unitary ensembles. We derive the  ladder operators satisfied by the associated monic orthogonal polynomials and  three compatibility conditions. By using them and introducing $2N$ auxiliary quantities $\{R_{n,j}, r_{n,j}, j=1,\cdots,N\}$, we build a series of difference equations.
Furthermore, we prove that $\{R_{n,j}, r_{n,j}\}$ satisfy  Riccati equations. From them we deduce a system of second order PDEs satisfied by $\{R_{n,j}, j=1,\cdots,N\}$, which  reduces to a Painlev\'{e} IV equation for $N=1$. We also show that the logarithmic derivative of the Hankel determinant  satisfies the generalized $\sigma$-form of a Painlev\'{e}  IV equation.\\
\textbf{Keywords}:  Gaussian unitary ensembles;  Hankel determinant; Orthogonal polynomials; Ladder operators; Painlev\'{e} equations\\
\textbf{Mathematics Subject Classification 2020}: 15B52; 33E17; 42C05
\end{abstract}
\noindent

\section{Introduction}

We consider the following average characteristic polynomial of $n\times n$ Hermitian matrices $H$ from the Gaussian unitary ensemble (GUE for short)
\[\langle\prod_{j=1}^N |\det (H-t_jI_n)|^{\gamma_j}\rangle_{{\rm GUE}}=\frac{D_n(\vec{t}\,)}{C_n},\]
where $I_n$ is the $n\times n$ identity matrix, $ t_j\in (-\infty ,+\infty )$ and $ \gamma _j> -1$ for $j=1,\cdots,N$, and $\vec{t}=(t_1,\cdots ,t_N)$. Here $D_n(\vec{t}\,)$ is given by
\begin{align*}
D_n(\vec{t}\,):=\frac{1}{n!}\int_{(-\infty, +\infty)^n }\prod_{1\le i< j\le n}(x_i-x_j)^2\prod_{k=1}^{n} w(x_k;\vec{t}\,)dx_1\cdots dx_n
\end{align*}
with
\begin{align}
	w(x;\vec{t}\, ):= {\rm e}^{-x^2} \prod_{j=1}^{N} \left | x-t_j \right | ^{\gamma _j}, \qquad x\in (-\infty ,+\infty ). \label{eq2}
\end{align}
The constant term $C_n$ has the following explicit expression \cite[p. 321]{17}
\begin{align*}
C_n:=&\frac{1}{n!}\int_{(-\infty, +\infty)^n }\prod_{1\le i< j\le n}(x_i-x_j)^2\prod_{k=1}^{n} \e^{-x_k^2}dx_1\cdots dx_n\\
=&(2\pi)^{n/2}2^{-n^2/2}\prod_{j=1}^{n-1}j!.
\end{align*}
The asymptotics of $\langle\prod\limits_{j=1}^N |\det (H-t_jI_n)|^{\gamma_j}\rangle_{{\rm GUE}}$ was established in \cite{Kra} when $\{t_j, j=1,\cdots,N\}$ tend to the bulk of the spectrum of $H$.

According to Heine's formula \cite{06}, one knows that $D_n(\vec{t} \,)$ can be evaluated as the determinant of the Hankel matrix generated by the moments of the weight function (\ref{eq2}), namely
\begin{align}
	D_n(\vec{t} \,)={\rm det}\left ( \int_{-\infty}^{+\infty }x^{i+j}w(x;\vec{t}\,)dx  \right ) _{i,j=0}^{n-1} .\label{eq1}
\end{align}
For the weight function $w(x;\vec{t}\,)$, we define the associated monic orthogonal polynomials by
\begin{align}
	\int_{-\infty }^{+\infty }P_n(x;\vec{t}\,)P_m(x;\vec{t}\,)w(x;\vec{t}\,)dx=h_n\delta _{mn}, \qquad m,n=0,1,2,\cdots, \label{eq3}
\end{align}
where $\delta_{mn}=1$ for $m=n$ and $0$ otherwise, and $P_n(x;\vec{t}\,)$ has the following form
\begin{align} P_n(x;\vec{t}\,):=x^n+p(n,\vec{t}\,)x^{n-1}+\cdots +P_n(0;\vec{t}\,).\label{eq4}
\end{align}
It is well-known that $D_n(\vec{t} \,)$ admits the following representation
\begin{align}
	D_n(\vec{t} \,)=\prod_{
		j=0}^{n-1} h_j(\vec{t} \,).\label{eq033}
\end{align}
In this paper, we will study $D_n(\vec{t} \,)$ by using the ladder operators satisfied by $\{P_n(x;\vec{t}\,), n=0,1,\cdots\}$ and three compatibility conditions. We call this the ladder operator approach and will describe it below.

We first look at the properties of $\{P_n(x;\vec{t}\,)\}$. From (\ref{eq3}) and (\ref{eq4}), one derives the following three-term recurrence relation
\begin{align}	 xP_n(x;\vec{t}\,)=P_{n+1}(x;\vec{t}\,)+\alpha_n(\vec{t}\,)P_n(x;\vec{t}\,)+\beta_n(\vec{t}\,)P_{n-1}(x;\vec{t}\,), \label{eq5}
\end{align}
with the initial conditions $P_0(x;\vec{t}\,):=1$ and $\beta _0P_{-1}(x;\vec{t}\,):=0$, and the recurrence coefficients are given by
\begin{align}
	\alpha _n(\vec{t}\,)&=p(n,\vec{t}\,)-p(n+1,\vec{t}\,), \qquad n\ge 0, \label{eq6}\\
	\beta _n(\vec{t}\,)&=\frac{h_n(\vec{t}\,)}{h_{n-1}(\vec{t}\,)}, \qquad n\ge 1, \label{eq7}
\end{align}
with $p(0,\vec{t}\,):=0$. It follows from (\ref{eq6}) that
\begin{align}
	 \sum_{j=0}^{n-1}\alpha_j(\vec{t}\,)=-p(n,\vec{t}\, ), \label{eq034}
\end{align}
and a combination of (\ref{eq033}) and (\ref{eq7}) gives us
$$\beta _n(\vec{t}\,)=\frac{D_{n+1}(\vec{t}\,) D_{n-1}(\vec{t}\,)}{D_n^2(\vec{t}\,)}.$$
Using the recurrence relation (\ref{eq5}), one obtains the standard Christoffel-Darboux formula
\begin{align} \sum_{k=0}^{n-1}\frac{P_k(x;\vec{t}\,)P_k(y;\vec{t}\,)}{h_k(\vec{t}\,)}= \frac{P_n(x;\vec{t}\,)P_{n-1}(y;\vec{t}\,)-P_n(y;\vec{t}\,)P_{n-1}(x;\vec{t}\,)}{h_{n-1}(\vec{t}\,)(x-y)}. \label{eq017}
\end{align}
With the above properties and the orthogonality relation \eqref{eq3}, one derives  a pair of ladder operators  satisfied by $P_n(z;\vec{t}\,)$ where two allied quantities $A_n(z)$ and $B_n(z)$ appear. Moreover, by using the ladder operators, one establishes three compatibility conditions for $A_n$ and $B_n$, numbered $(S_1),(S_2)$ and $(S_2')$, which formulates the ladder operator method.
See \cite{06,18} for a detailed description.

The special case where the weight function (\ref{eq2}) has only one  discontinuity (i.e. $N=1$) was studied in \cite{02}. Through the ladder operator approach, the recurrence coefficient $\alpha_n$ was shown to satisfy a particular Painlev\'{e} IV equation.
It should be pointed out that our Hankel determinant $D_n(\vec{t}\,)$ given by (\ref{eq1}) was raised in \cite{02} with few further discussions.
In this paper, although we will still use the ladder operator approach to investigate $D_n(\vec{t}\,)$ with $N$ singularities, the derivation and results are not simple generalization of those in \cite{02} where $N=1$. Our main contribution is as follows.

In \cite{02}, the authors mainly made use of the asymptotics for $A_n(z)$ and $B_n(z)$ as $z\rightarrow+\infty$ to obtain the Painlev\'{e} IV equation for $\alpha_n$. However, this derivation strategy is not applicable to our problem. To derive PDEs to characterize $D_n(\vec{t}\,)$, we have to separate each integral that appears in $A_n(z)$ and $B_n(z)$ into two parts, in one of which we introduce auxiliary quantities $\{R_{n,j},r_{n,j}\}$ (see \eqref{eq23} and \eqref{eq24}).
We give a detailed explanation regarding the above discussion in Section \ref{explain}.

In addition, to obtain the PDEs satisfied by $\{R_{n,j}\}$, we make use of the compatibility conditions $\frac{\partial}{\partial t_k}R_{n,j}=\frac{\partial}{\partial t_j}R_{n,k}$ and $\frac{\partial}{\partial t_k}r_{n,j}=\frac{\partial}{\partial t_j}r_{n,k}$ for $j,k=1,\cdots,N$ which are derived from the differentiation of the orthogonality relation. When $N=1$, these PDEs for $\{R_{n,j}\}$ are reduced to the Painlev\'{e} IV equation of \cite{02}.
Moreover, in our manuscript we obtain the PDE satisfied by $\sigma_n$, the logarithmic derivative of our Hankel determinant, which was not established in \cite{02} for $N=1$. In order to derive the desired PDE, we substitute the asymptotics of $A_n(z)$ and $B_n(z)$ as $z\rightarrow+\infty$ into $(S_2')$ to obtain the crucial identity which expresses $\sigma_n$ in terms of the auxiliary quantities $\{R_{n,j},r_{n,j}\}$. In turn, we express $\{R_{n,j},r_{n,j}\}$ in terms of $\sigma_n$ and its first order derivatives, where the expressions for $\{R_{n,j}\}$ are obtained from a quadratic equation which comes from the Riccati equation for $r_{n,j}$ and the existence of real solutions of which is discussed. Finally, we come to the second order nonlinear PDE for $\sigma_n$ which seems simple in form and is reduced to the $\sigma$-form of a Painlev\'{e} IV equation for $N=1$.

The ladder operator approach has been widely used in the study of orthogonal polynomials and random matrix ensembles. For example, it was taken to derive the properties of classical monic Jacobi polynomials \cite{08} including their recurrence coefficients, the square of their $L^2$-norms and their explicit representations.
In \cite{09}, the Hankel determinant for the weight function $x^\alpha {\rm e}^{-x-s/x}, x\in \left [ 0,\infty \right )  ,\alpha>0,s>0$ was studied, which arises from  an integrable quantum field theory at finite temperature. The ladder operator approach as well as the Lax pair of the Riemann-Hilbert problem for the associated orthogonal polynomials were used to derive the integral representation for the Hankel determinant in terms of solutions of a  Painlev\'{e} \uppercase\expandafter{\romannumeral3} equation.
The ladder operator approach was also adopted to study unitary ensembles with the weight function having two or more variables, and a second order PDE was established for the logarithmic derivative of the corresponding Hankel determinant; see e.g. \cite{10,11,13,12,15,16}.

Random matrix ensembles with Fisher-Hartwig singularities of both jump and root types at the same point have attracted extensive attention recently. In \cite{03}, a Fisher-Hartwig singularity of jump type was added to the weight function of \cite{02}. By using the ladder operator approach, a Painlev\'{e} IV equation was established for finite $n$, and as $n\to \infty$, the asymptotics of the recurrence coefficients and the Hankel determinant were obtained in terms of solutions of the $\sigma $-form of a Painlev\'{e} XXXIV equation at the hard edge and of a Painlev\'{e} II equation at the soft edge respectively.
The Hankel determinant generated by the perturbed Laguerre weight $x^\alpha {\rm e}^{-x}\left | x-t \right |^\gamma (A+B\theta (x-t))$,  $x,t\ge 0$, $\alpha ,\gamma > 0$ was investigated in \cite{04}. By taking the ladder operator approach, the logarithmic derivative of the Hankel determinant was found to satisfy  the $\sigma $-form of a Painlev\'{e} \uppercase\expandafter{\romannumeral5} equation. And by adopting Dyson's Coulomb fluid method,
the asymptotic behavior of the Hankel determinant at the soft edge is characterized by a Painlev\'{e} \uppercase\expandafter{\romannumeral34} equation.
The Hankel determinant generated by the Jacobi weight
$x^\alpha(1-x)^\beta \left | x-t \right |^\gamma (A+B\theta (x-t)) $, $x,t\in \left [ 0,1 \right ] $, $\alpha ,\beta ,\gamma > 0$ was studied in \cite{05}.  Via the ladder operator approach, the logarithmic derivative of the Hankel determinant was shown to satisfy the $\sigma $-form of a Painlev\'{e} \uppercase\expandafter{\romannumeral6} equation for finite dimension $n$ and of a Painlev\'{e} \uppercase\expandafter{\romannumeral3} equation under suitable double scaling.

The gap probability of the circular unitary ensemble with a Fisher-Hartwig singularity of both jump and root types was studied in \cite{07}. By employing Deift-Zhou nonlinear steepest descent analysis to the Riemann-Hilbert (RH for short) problem satisfied by the associated orthogonal polynomials (known as the RH method) to study the asymptotics of the Toeplitz determinant, the gap probability of the circular unitary ensemble was represented as an integral of the Hamiltonian of a coupled Painlev\'{e} \uppercase\expandafter{\romannumeral5} system. The Hankel determinant for the Laguerre weight as well as the Jacobi weight with several Fisher-Hartwig singularities of both root type and jump type at the same point were investigated in \cite{21} by using the RH method, and the asymptotics for the Hankel determinants were obtained at the soft or hard edge. See also \cite{01}. The RH method was used widely to study unitary ensembles; see e.g. \cite{22,21,24,23,20}.

In this paper, we take the ladder operator approach to study the Hankel determinant given by \eqref{eq1}, and its logarithmic derivative is shown to satisfy the generalized $\sigma$-form of a Painlev\'{e} IV equation.
The outline is as follows. In Section $2$,  we derive the ladder operators and compatibility conditions for the monic orthogonal polynomials defined by \eqref{eq3} and \eqref{eq4}.
By using them, we obtain in Section $3$ a series of difference equations for the auxiliary quantities  $\{R_{n,j}, r_{n,j}, j=1,\cdots,N\}$ introduced in the ladder operators.
In Section 4, we deduce Toda equations for the recurrence coefficients and Riccati equations satisfied by $\{R_{n,j}, r_{n,j}\}$ from which a system of second order PDEs are obtained for $\{R_{n,j}\}$. When $N=1$, these PDEs are reduced to a particular
Painlev\'{e} IV equation. Based on the above results, we finally establish in Section $5$ a second order PDE satisfied by the logarithmic derivative of the Hankel determinant, which is reduced to the $\sigma$-form of a Painlev\'{e} IV equation when $N=1$.

\section{Ladder operators and compatibility conditions}
In this section, we use the definition and properties of orthogonal polynomials to derive lowering and raising operators and three compatibility conditions ($S_1$), ($S_2$), ($S_2'$).

Before concentrating on our problem, we consider a more general case. Denote
\begin{align}
w(x;\vec{t}\,)=&w_0(x)w_F(x;\vec{t}\,),\qquad x\in[c,d],\label{w-1}\\
w_F(x;\vec{t}\,):=&\prod_{j=1}^{N} \left | x-t_j \right | ^{\gamma _j},\qquad  t_j\in [c,d], \,\gamma_j>-1, \, j=1,\cdots ,N,\nonumber
\end{align}
where $w_0(x)$ is an arbitrary positive smooth function on $[c,d]$ whose moments of all orders exist, and $w_0(c)=w_0(d)=0$.
For our problem, $w_0(x)=\e^{-x^2}, x\in(-\infty,+\infty)$.
In the following discussions, we shall not display the  dependence of $\vec{t}\,$ when not necessary.

\begin{theorem} The monic polynomials $\{P_n(z),n=0,1,\cdots\}$ orthogonal with respect to \eqref{w-1} satisfy the following lowering and raising operators
\begin{align}
		\biggl (\frac{d}{dz}+B_n(z) \biggr)P_n(z)=\beta_nA_n(z)P_{n-1}(z),\label{eq8}
\end{align}
\begin{align}
\biggl (\frac{d}{dz}-B_n(z)-{\rm v}_0'(z) \biggr )P_{n-1}(z)=-A_{n-1}(z)P_n(z),\label{eq9}
\end{align}
where $A_n(z)$ and $B_n(z)$ are defined by
\begin{align}
	A_n(z)&:=\frac{1}{h_n} \int_{c}^{d} \frac{{\rm v}_0'(z)-{\rm v}_0'(y)}{z-y} P_{n}^{2}(y)w(y)dy+a_{n}(z),\label{eq10}\\
	B_n(z)&:=\frac{1}{h_{n-1}} \int_{c}^{d} \frac{{\rm v}_0'(z)-{\rm v}_0'(y)}{z-y} P_{n}(y)P_{n-1}(y)w(y)dy+b_{n}(z).\label{eq11}
\end{align}
Here ${\rm v}_0(x)=-\ln{w_0(x)}$ and $\{a_{n}(z), b_{n}(z)\}$ read
\begin{equation}\label{defanbn}
\begin{aligned}
a_{n}(z):=&\sum_{j=1}^{N}\frac{\gamma _j}{h_n} \int_{c}^{d}  \frac{ P_{n}^2(y)w(y)}{(y-t_j)(z-y)}dy,\\
b_{n}(z):=&\sum_{j=1}^{N}\frac{\gamma _j}{h_{n-1}}  \int_{c}^{d}  \frac{ P_{n}(y)P_{n-1}(y)w(y) }{(y-t_j)(z-y)}dy.
\end{aligned}
\end{equation}
\end{theorem}
\begin{proof}
	Since $P_n(z) $ is a polynomial of degree $n$, we have
	\begin{align}
		P_{n}'(z)=\sum_{k=0}^{n-1} C_{n,k}P_k(z),\label{eq12}
	\end{align}
where, according to the orthogonality relation (\ref{eq3}),  the coefficients are given by $$C_{n,k}=\frac{1}{h_k} \int_{c}^{d} P_{n}'(y)P_k(y)w(y)dy, \qquad k=0,1,\cdots ,n-1.$$
Inserting the above expression back into (\ref{eq12}), through integration by parts and in view of $w(c)=w(d)=0$, we find
	\begin{align}
	P_{n}'(z)=&\sum_{k=0}^{n-1} \frac{P_k(z)}{h_k}\int_{c}^{d}P_k(y)w(y)d(P_n(y))\nonumber\\
	=& -\sum_{k=0}^{n-1} \frac{P_k(z)}{h_k}\left[ \int_{c}^{d}P_n(y)P_{k}'(y)w(y)dy +  \int_{c}^{d}P_n(y)P_k(y)w_{0}'(y)w_F(y)dy\right.\nonumber\\
	&\qquad \qquad \qquad \left.+ \int_{c}^{d}P_n(y)P_k(y)w_{0}(y)w_F'(y)dy\right].\label{eq13}
	\end{align}

We now look at the three integrals in the above square bracket one by one.
Since for $k=1,\cdots ,n-1$, $P_k'(y)$ is an orthogonal polynomial of degree $k-1$, which is at most $n-2$, we know from the orthogonality relation (\ref{eq3}) that the first term in the square bracket is zero.
For the second integral, according to the identity
$w_0'(y)=-{\rm v}_0'(y)w_0(y)$ and the orthogonality relation (\ref{eq3}), it follows that
\begin{align}
	 \int_{c}^{d}P_n(y)P_k(y)w_{0}'(y)w_F(y)dy=\int_{c}^{d}P_n(y)P_k(y)({\rm v}_0'(z)-{\rm v}_0'(y))w(y)dy,\label{eq14}
\end{align}
for $k=0,1,\cdots ,n-1$.
To study the third integral, we need the following facts
$$|y-t_j|^{\gamma_j}=(y-t_j)^{\gamma_j} \theta (y-t_j)+(t_j-y)^{\gamma_j} \theta (t_j-y),$$
$$\frac{d}{dx}\theta (x)=\delta (x), $$
where $\theta(x)$ is the Heaviside step function which is 1 for $x>0$ and 0 otherwise, and $\delta(\cdot)$ is the Dirac delta function. From the above two identities, it follows that
$$\frac{\partial }{\partial y}|y-t_j|^{\gamma_j}=\delta (y-t_j)((y-t_j)^{\gamma_j}-(t_j-y)^{\gamma_j} )+\gamma_j \frac{|y-t_j|^{\gamma_j}}{y-t_j}.$$
Hence, the third integral in \eqref{eq13} now reads
\begin{align}
\int_{c}^{d}P_n(y)P_k(y)w_0(y)w_{F}'(y)dy=& \int_{c }^{d}P_n(y)P_k(y)w_0(y)\left(\frac{\partial }{\partial y}|y-t_1|^{\gamma _1}\cdots |y-t_N|^{\gamma _N}+\right.\nonumber
	\\&\qquad\left.\cdots+|y-t_1|^{\gamma _1}\cdots\frac{\partial }{\partial y}|y-t_N|^{\gamma_N}\right)dy \nonumber
	 \\=&\int_{c}^{d}P_n(y)P_k(y)\sum_{j=1}^{N}\frac{\gamma _j}{y-t_j}w(y)dy.\label{eq15}
\end{align}
Plugging (\ref{eq14}) and (\ref{eq15}) into (\ref{eq13}), we get
\begin{align*}
	P_{n}^{'}(z)=&-\int_{c}^{d} \sum_{k=0}^{n-1}\frac{P_k(z)P_k(y)}{h_k}P_n(y)\biggl[({\rm v}_0'(z)-{\rm v}_0'(y))w(y)+\sum_{j=1}^{N}\frac{\gamma _j}{y-t_j}w(y)\biggr]dy \nonumber\\
	=&-\int_{c}^{d} \frac{P_n(z)P_{n-1}(y)-P_n(y)P_{n-1}(z)}{h_{n-1}(z-y)}P_n(y)\biggl[({\rm v}_0'(z)-{\rm v}_0'(y))w(y)+\sum_{j=1}^{N}\frac{\gamma _j}{y-t_j}w(y)\biggr]dy \nonumber\\
	=&-\frac{P_n(z)}{h_{n-1}}\biggl ( \int_{c}^{d} \frac{{\rm v}_0'(z)-{\rm v}_0'(y)}{z-y} P_{n}(y)P_{n-1}(y)w(y)dy+\sum_{j=1}^{N}\gamma _j \int_{c}^{d}  \frac{ P_{n}(y)P_{n-1}(y)w(y) }{(y-t_j)(z-y)}dy \biggr ) \nonumber\\
	&+\frac{P_{n-1}(z)}{h_{n-1}}\biggl (  \int_{c}^{d} \frac{{\rm v}_0'(z)-{\rm v}_0'(y)}{z-y} P_{n}^2(y)w(y)dy+\sum_{j=1}^{N}\gamma _j \int_{c}^{d}  \frac{ P_{n}^2(y)w(y)}{(y-t_j)(z-y)}dy \biggr ) \nonumber\\
	=&-B_n(z)P_n(z)+\beta_n A_n(z)P_{n-1}(z), \label{eq16}
\end{align*}
where the second equality is due to the Christoffel-Darboux formula \eqref{eq017} and to get the last identity we make use of the fact that $\beta _n=h_n /h_{n-1}$.
This completes the proof of the lowering operator (\ref{eq8}).

Replacing $n$ by $n-1$ in (\ref{eq8}) and the recurrence relation (\ref{eq5}), we have
\begin{align*}
P_{n-1}'(z)=&\beta _{n-1}A_{n-1}(z)P_{n-2}(z)-B_{n-1}(z)P_{n-1}(z),\\
\beta _{n-1}P_{n-2}(z)=&(z-\alpha _{n-1})P_{n-1}(z)-P_{n}(z).
\end{align*}
Substituting the second equation into the first one,
we get
$$P_{n-1}'(z)=\left[(z-\alpha _{n-1})A_{n-1}(z)-B_{n-1}(z)\right]P_{n-1}(z)-A_{n-1}(z)P_{n}(z).$$
According to (\ref{$S_1$}) which will be given and proved in the next theorem, we know that the term in the above square bracket is $B_n(z)+{\rm v}_0'(z)$. Hence, we are led to the raising operator (\ref{eq9}).
\end{proof}
\begin{remark}
\eqref{eq10}-\eqref{defanbn} with $c=-\infty$ and $d=+\infty$ were given by $(3.9)$-$(3.11)$ of \cite{02} without a proof.
\end{remark}

\begin{theorem} The functions $A_n(z)$ and $B_n(z)$ satisfy the equations
	\begin{equation}
		B_{n+1}(z)+B_n(z)=(z-\alpha _n)A_n(z)-{\rm v}_0'(z),\tag{$S_1$}\label{$S_1$}
	\end{equation}
	\begin{equation}
		1+(z-\alpha _n)(B_{n+1}(z)-B_n(z))=\beta _{n+1}A_{n+1}(z)- \beta _{n}A_{n-1}(z).\tag{$S_2$}\label{$S_2$}
	\end{equation}
\end{theorem}
\begin{proof}
According to the definition of $B_n(z)$ given by \eqref{eq11}, we have
	\begin{equation}		
		\begin{split}	 B_{n+1}(z)+B_{n}(z)=&\int_{c}^{d}\frac{{\rm v}_0'(z)-{\rm v}_0'(y)}{z-y}\left(\frac{P_{n+1}(y)}{h_n}+\frac{P_{n-1}(y)}{h_{n-1}}\right)P_n(y)w(y)dy\\
	&+\sum_{j=1}^{N}\gamma _j\int_{c}^{d} \left(\frac{P_{n+1}(y)}{h_n}+\frac{P_{n-1}(y)}{h_{n-1}}\right)\frac{P_n(y)w(y)}{(z-y)(y-t_j)}dy.\label{eq18}	 \end{split}		
	\end{equation}
Since it follows from the recurrence relation (\ref{eq5}) and $\beta _n={h_n}/{h_{n-1}}$ that
$$\frac{P_{n+1}(y)}{h_n}+\frac{P_{n-1}(y)}{h_{n-1}}=\frac{(y-\alpha _n)P_n(y)}{h_n},$$
by substituting it into (\ref{eq18}), we come to		 
	\begin{align}
		 &B_{n+1}(z)+B_{n}(z)\nonumber\\
&\;=\frac{1}{h_n} \int_{c}^{d}\frac{{\rm v}_0'(z)-{\rm v}_0'(y)}{z-y}(y-\alpha _n)P_n^2(y)w(y)dy+\sum_{j=1}^{N}\frac{\gamma_j }{h_n}\int_{c}^{d}(y-\alpha _n)\frac{P_n^2(y)w(y)dy}{(z-y)(y-t_j)}
\nonumber\\
&\;=(z-\alpha _n)\biggl(\frac{1}{h_{n}}  \int_{c}^{d} \frac{{\rm v}_0'(z)-{\rm v}_0'(y)}{z-y} P_{n}^2(y)w(y)dy+\sum_{j=1}^{N}\frac{\gamma _j}{h_{n}} \int_{c}^{d}  \frac{ P_{n}^2(y)w(y)}{(y-t_j)(z-y)}dy\biggr)\nonumber\\
		&\qquad-\frac{1}{h_n} \int_{c}^{d}({\rm v}_0'(z)-{\rm v}_0'(y))P_n^2(y)w(y)dy-\sum_{j=1}^{N}\frac{\gamma_j }{h_n}\int_{c}^{d }P_n^2(y)\frac{w(y)}{y-t_j}dy\nonumber\\
&=\;(z-\alpha_n)A_n(z)-{\rm v}_0'(z)+\frac{1}{h_n} \int_{c}^{d}P_n^2(y){\rm v}_0'(y)w(y)dy-\sum_{j=1}^{N}\frac{\gamma_j }{h_n}\int_{c}^{d }P_n^2(y)\frac{w(y)}{y-t_j}dy,\label{Bn+1}
\end{align}
where to get the second equality we make use of the identity
$$\frac{y-\alpha _n}{z-y} =\frac{z-\alpha _n}{z-y}-1,$$
and the last equation is obtained by using the definition of $A_n(z)$ given by \eqref{eq10} and the orthogonality condition (\ref{eq3}).
Now we look at the first integral on the right hand side of \eqref{Bn+1}. Noting that ${\rm v}_0'(y)w(y)={\rm v}_0'(y)w_0(y)w_F(y)=-w_0'(y)w_F(y)$, through  integration by parts and in view of $w(c)=w(d)=0$, we find
\begin{align}
	\frac{1}{h_n}\int_{c}^{d}P_n^2(y){\rm v}_0'(y)w(y)dy=&-\frac{1}{h_n} \int_{c}^{d}P_n^2(y)w_{0}'(y)w_F(y)dy\nonumber\\
=&\frac{1}{h_n} \int_{c }^{d}2P_n(y)P_n'(y)w(y)dy + \frac{1}{h_n}\int_{c }^{d }P_n^2(y)w_0(y)w_F'(y)dy.\label{eq21-1}
\end{align}
Since $P_n'(y)$ is a monic orthogonal polynomial of degree $n-1$, according to the orthogonal relation \eqref{eq3}, we know that the first integral in \eqref{eq21-1} is zero. Via an argument similar to the derivation of (\ref{eq15}) to deal with the second integral in \eqref{eq21-1}, we obtain
\begin{align}
	\frac{1}{h_n}\int_{c}^{d}P_n^2(y){\rm v}_0'(y)w(y)dy=\sum_{j=1}^{N} \frac{\gamma_j}{h_n} \int_{c}^{d }P_n^2(y)\frac{w(y)}{y-t_j}dy.\label{eq21}
\end{align}
Plugging it back into \eqref{Bn+1}, we arrive at (\ref{$S_1$}).

We now turn to the derivation of (\ref{$S_2$}). Replacing $x$ by $z$ in the recurrence relation (\ref{eq5}) and differentiating it with respect to $z$, we get
\begin{align}
P_{n+1}'(z)=(z-\alpha _n)P_{n}'(z)+P_{n}(z)-\beta _nP_{n-1}'(z) .\label{eq22}
\end{align}
Replacing $n$ by $n+1$ in the lowering operator (\ref{eq8}), and rewriting (\ref{eq8}) and (\ref{eq9}), we have
\begin{align}
	P_{n+1}'(z)&=\beta _{n+1}A_{n+1}(z)P_{n}(z)-B_{n+1}(z)P_{n+1}(z),\nonumber\\
	P_{n}'(z)&=\beta _{n}A_{n}(z)P_{n-1}(z)-B_{n}(z)P_{n}(z),\nonumber\\
	P_{n-1}'(z)&=\left(B_n(z)+{\rm v}_0'(z)\right)P_{n-1}(z)-A_{n-1}(z)P_{n}(z).\nonumber
\end{align}
Substituting  them into (\ref{eq22}) and eliminating $P_{n+1}(z)$ in the resulting equation by using the three-term recurrence relation (\ref{eq5}), we are led to
\begin{equation}		
	\begin{split}
		&[\beta _{n+1}A_{n+1}(z)- \beta _{n}A_{n-1}(z)-(z-\alpha _n)(B_{n+1}(z)-B_n(z))-1]P_n(z)\\
		=&\beta _n[(z-\alpha _n)A_n(z)-B_{n+1}(z)-B_n(z)-{\rm v}_0'(z)]P_{n-1}(z).\nonumber
	\end{split}		
\end{equation}
 According to (\ref{$S_1$}), we know that the right hand side of the above equation is zero. Hence it follows that
$$\beta _{n+1}A_{n+1}(z)- \beta _{n}A_{n-1}(z)-(z-\alpha _n)(B_{n+1}(z)-B_n(z))-1=0,$$
which completes the proof of (\ref{$S_2$}).
\end{proof}
The combination of (\ref{$S_1$}) and (\ref{$S_2$}) produces a sum rule.
\begin{theorem}  $A_n(z)$ and $B_n(z)$ satisfy the following equation
  \begin{equation}
	B_{n}^{2}(z)+{\rm v}_0'(z)B_n(z)+\sum_{j=0}^{n-1}A_j(z)=\beta _nA_n(z)A_{n-1}(z).\tag{$S_{2}'$}\label{$S_{2}'$}
  \end{equation}
\end{theorem}
\begin{proof}
Multiplying both sides of (\ref{$S_2$}) by $A_n(z)$ and replacing the term $(z-\alpha _n)A_n(z)$ in the resulting identity by $B_{n+1}(z)+B_n(z)+{\rm v}_0'(z)$, which is due to (\ref{$S_1$}), we find
$$A_n(z)+B_{n+1}^2(z)-B_{n}^2(z)+{\rm v}_0'(z)(B_{n+1}(z)-B_{n}(z))=\beta _{n+1}A_{n+1}(z)A_{n}(z)-\beta _{n}A_{n}(z)A_{n-1}(z).$$
Replacing $n$ by $j$ in this equality and summing over $j$ from $0$ to $n-1$, noting that $A_{-1}(z)=B_0(z)=0$, we arrive at (\ref{$S_{2}'$}).
\end{proof}

\begin{remark}
The derivation of $(S_2)$ by using $(S_1)$ and the ladder operators and of $(S_2')$ by combining $(S_1)$ and $(S_2)$ were presented in \cite{18}. See also \cite{ChenIsmail} and \cite{15}.
\end{remark}

\section{Difference equations}

In this section, we make use of the three compatibility conditions (\ref{$S_1$}), (\ref{$S_2$}) and (\ref{$S_{2}'$}) to express the recurrence coefficients and the coefficient of $x^{n-1}$ in the monic orthogonal polynomial $P_n(x;\vec{t}\,)$ in terms of the auxiliary quantities which will be introduced later and shown to satisfy a system of difference equations that can be iterated in $n$.

Before proceeding further, we first explain why the analysis presented in \cite{02} for the Hankel determinant \eqref{eq1} with $N=1$ can not be generalized to our problem with $N$ generic.
\subsection{Comparison between our problem and the $N=1$ case}\label{explain}
Comparing our weight function \eqref{eq2} with \eqref{w-1}, we know that $w_0(x)={\rm e}^{-x^2},~c=-\infty,~d=+\infty$. Hence ${\rm v}_0(x)=x^2$, so that
\[ \frac{{\rm v}_0'(z)-{\rm v}_0'(y)}{z-y} =2.\]
Inserting it into (\ref{eq10}) and (\ref{eq11}), we readily get
\begin{align}
A_n(z)=&2+a_n(z),\label{Anan}\\
B_n(z)=&b_n(z),\label{Bnbn}
\end{align}
where $a_n(z)$ and $b_n(z)$ are given by
\begin{equation}\label{anbn-N}
\begin{aligned}
a_{n}(z):=&\sum_{j=1}^{N}\frac{\gamma _j}{h_n} \int_{-\infty}^{+\infty}  \frac{ P_{n}^2(y)w(y)}{(y-t_j)(z-y)}dy,\\
b_{n}(z):=&\sum_{j=1}^{N}\frac{\gamma _j}{h_{n-1}}  \int_{-\infty}^{+\infty}  \frac{ P_{n}(y)P_{n-1}(y)w(y) }{(y-t_j)(z-y)}dy.
\end{aligned}
\end{equation}

The special case of our problem \eqref{eq1} with $N=1$, i.e. the Hankel determinant for the weight function $w(x,t)={\e}^{-x^2}|x-t|^{\gamma}$, was studied in \cite{02}. By inserting
\begin{align}	 \frac{1}{z-y}=\frac{1}{z}+\frac{y}{z^2}+\frac{y^2}{z^3}+O\biggl(\frac{1}{z^4}\biggr),\qquad z\to +\infty, \label{eq043}
\end{align}
into \eqref{Anan}-\eqref{anbn-N} with $N=1$, the asymptotic expansions of $A_n(z)$ and $B_n(z)$ were obtained in \cite{02}:
\begin{equation}\label{AnBn-1}
\begin{aligned}
A_n(z)&\sim 2+\frac{2\alpha_n}{z}+\frac{\gamma+2t\alpha_n}{z^2}+\frac{\gamma t+\gamma \alpha_n+2t^2\alpha_n}{z^3}+\cdots,\\
B_n(z)&\sim\frac{2\beta_n-n}{z}+\frac{t(2\beta_n-n)}{z^2}+\frac{\gamma\beta_n+t^2(2\beta_n-n)}{z^3}+\cdots.
\end{aligned}
\end{equation}
Since the coefficients in the above asymptotics are intimately related to $\alpha_n$ and $\beta_n$, by plugging \eqref{AnBn-1} into the compatibility conditions $(S_1)$ and $(S_2)$, a coupled difference equations were established for $\alpha_n$ and $\bt_n$ (see (4.5) and (4.6) of \cite{02}), which are crucial for the derivation of the Painlev\'{e} IV equation satisfed by $\alpha_n$.

 Motivated by this idea, we substitute \eqref{eq043} into \eqref{Anan}-\eqref{anbn-N} to derive the asymptotics for $A_n$ and $B_n$ as $z\rightarrow+\infty$. We get
\begin{align}
	 A_{n}(z)=&2+\frac{1}{z}\sum_{j=1}^{N}\frac{\gamma _j}{h_n} \int_{-\infty }^{+\infty }  \frac{ P_{n}^2(y)w(y)}{y-t_j}dy +\frac{1}{z^2}\sum_{j=1}^{N}\frac{\gamma _j}{h_n} \int_{-\infty }^{+\infty } \biggl(1+ \frac{t_j}{y-t_j}\biggr)P_{n}^2(y)w(y)dy\nonumber\\
	&\;\,+\frac{1}{z^3}\sum_{j=1}^{N}\frac{\gamma _j}{h_n} \int_{-\infty }^{+\infty }\biggl(y+t_j+ \frac{t_j^2}{y-t_j}\biggr)P_{n}^2(y)w(y)dy+O\biggl(\frac{1}{z^4}\biggr)\nonumber\\	 =&2+\frac{\sum\limits_{j=1}^{N}R_{n,j}}{z}+\frac{\sum\limits _{j=1}^{N}\gamma _j+\sum\limits _{j=1}^{N}t_jR_{n,j}}{z^2}+\frac{\sum\limits _{j=1}^{N}\gamma _j(\alpha _n+t_j)+\sum\limits _{j=1}^{N}t_j^2R_{n,j}}{z^3} +O\biggl(\frac{1}{z^4}\biggr),\label{An-asy}
\end{align}
where the auxiliary quantities $\{R_{n,j}, j=1,\cdots,N\}$ are defined by
\begin{align}\label{defRn-1}
	R_{n,j}(\vec{t}\, ):=\frac{\gamma_j }{h_n}\int_{-\infty }^{+\infty } \frac{P_{n}^{2}(y) }{y-t_j}w(y)dy,
\end{align}
and to get \eqref{An-asy} we make use of the orthogonality relation (\ref{eq3}) and the recurrence relation (\ref{eq5}). Similarly, we can show that $B_n(z)$ has the following asymptotics as $z\rightarrow+\infty:$
\begin{align}	
B_{n}(z)=&\frac{\sum\limits_{j=1}^{N}r_{n,j}}{z}+\frac{\sum\limits _{j=1}^{N}t_jr_{n,j}}{z^2}+\frac{\beta _n\sum\limits _{j=1}^{N}\gamma _j+\sum\limits _{j=1}^{N}t_j^2r_{n,j}}{z^3}+O\biggl(\frac{1}{z^4}\biggr).\label{Bn-asy}
\end{align}
where the auxiliary quantities $\{r_{n,j}, j=1,\cdots,N\}$ are given by
\begin{align}\label{defrn-1}
	r_{n,j}(\vec{t}\, ):=\frac{\gamma_j }{h_{n-1}}\int_{-\infty }^{+\infty } \frac{P_{n}(y)P_{n-1}(y) }{y-t_j}w(y)dy.
\end{align}
Inserting ${\rm v}_0'(y)=2y, c=-\infty, d=+\infty$ into both sides of \eqref{eq21}, in view of the recurrence relation \eqref{eq5}, we find
\begin{align}\label{Ral}
\sum_{j=1}^{N}R_{n,j}=2\alpha _n.
\end{align}
Via an argument similar to the derivation of (\ref{eq21}), we get
\begin{align*}
	 \frac{1}{h_{n-1}}\int_{-\infty}^{+\infty}P_n(y)P_{n-1}(y){\rm v}_0'(y)w(y)dy
=&n+\sum_{j=1}^{N}r_{n,j}.
\end{align*}
Replacing ${\rm v}_0'(y)$ by $2y$ in the above identity, with the aid of the recurrence relation \eqref{eq5}, we are led to
\begin{align}\label{rbt}
\sum_{j=1}^{N}r_{n,j}=2\beta _n-n.
\end{align}
Plugging \eqref{Ral} and \eqref{rbt} into \eqref{An-asy} and \eqref{Bn-asy}, we have
\begin{equation}\label{AnBn-N}
\begin{aligned}
A_n(z)=&2+\frac{2\alpha_n}{z}+\frac{\sum\limits _{j=1}^{N}\gamma _j+\sum\limits _{j=1}^{N}t_jR_{n,j}}{z^2}+\frac{\sum\limits _{j=1}^{N}\gamma _j(\alpha _n+t_j)+\sum\limits _{j=1}^{N}t_j^2R_{n,j}}{z^3} +O\biggl(\frac{1}{z^4}\biggr),\\
B_n(z)=&\frac{2\bt_n-n}{z}+\frac{\sum\limits _{j=1}^{N}t_jr_{n,j}}{z^2}+\frac{\beta _n\sum\limits _{j=1}^{N}\gamma _j+\sum\limits _{j=1}^{N}t_j^2r_{n,j}}{z^3}+O\biggl(\frac{1}{z^4}\biggr).
\end{aligned}
\end{equation}
We observe that the coefficients of $z^{-2}$ and $z^{-3}$ involve several summation terms, which is different from the $N=1$ case where the coefficients are uniquely determined by $\alpha_n$ and $\beta_n$ (see \eqref{AnBn-1}). Hence, by substituting \eqref{AnBn-N} into $(S_1)$ and $(S_2)$, we are unable to deduce for $\alpha_n$ and $\beta_n$ the coupled difference equations to derive the Painlev\'{e} IV equation for $\alpha_n$ as was done in \cite{02}.

Actually, by inserting \eqref{An-asy} and \eqref{Bn-asy} into $(S_1)$ and comparing its both sides the coefficients of $z^{-1}$ and $z^{-2}$, we get
\begin{align}
\sum_{j=1}^N  (r_{n+1,j}+r_{n,j})=\sum_{j=1}^N\left((t_j-\alpha_n)R_{n,j}+\gamma_j\right).
\end{align}
Similarly, by using $(S_2)$, we find
\begin{align}
\sum_{j=1}^N
(t_j-\alpha_n)(r_{n+1,j}+r_{n,j})&=\sum_{j=1}^N(\beta_{n+1}R_{n+1,j}-\beta_n R_{n-1,j}).
\end{align}
If we continue to look at the coefficients of $z^{-k}$ for $k\geq3$ on both sides of $(S_1)$ and $(S_2)$, we find that the calculation is complicated and the identities obtained are also related to summation terms involving $\{t_j,R_{n,j},r_{n,j}\}$. With these equalities, we are not able to establish equations for $\{R_{n,j},r_{n,j}\}$ with $j$ given. Consequently, the derivation technique employed in \cite{02} for $N=1$ is not applicable to our problem where $N$ is generic. We have to use a different strategy to derive the desired equation to characterize the Hankel determinant \eqref{eq1}. We will see in the subsequent discussions that the quantities defined by \eqref{defRn-1} and \eqref{defrn-1} play an essential role.

\subsection{Difference equations for auxiliary quantities}
Note that
$$ \frac{1}{(z-y)(y-t_j)}=\frac{1}{z-t_j} \biggl(\frac{1}{y-t_j}+\frac{1}{z-y}\biggr).$$
Plugging it into \eqref{Anan}-\eqref{anbn-N}, we come to the following expressions for
$A_n(z)$ and $B_n(z)$.
\begin{lemma}
$A_n(z)$ and $B_n(z)$ are given by
	\begin{align}
		A_n(z)&:= 2+\sum_{j=1}^{N}\frac{R_{n,j} }{z-t_j}+\sum_{j=1}^{N}\frac{\gamma _j}{h_n(z-t_j) } \int_{-\infty }^{+\infty }  \frac{P_{n}^{2}(y) }{z-y}w(y)dy,\label{eq23}\\
		B_n(z)&:=\sum_{j=1}^{N}\frac{r_{n,j} }{z-t_j}+\sum_{j=1}^{N}\frac{\gamma _j}{h_{n-1}(z-t_j) } \int_{-\infty }^{+\infty }  \frac{ P_{n}(y)P_{n-1}(y) }{z-y}w(y)dy,\label{eq24}
	\end{align}
where the auxiliary quantities $\{R_{n,j}, r_{n,j}, j=1,\cdots,N\}$ are defined by \eqref{defRn-1} and \eqref{defrn-1}.
\end{lemma}
Substituting (\ref{eq23}) and (\ref{eq24}) into (\ref{$S_1$}) and (\ref{$S_2$}), we obtain several difference equations and expressions.
From (\ref{$S_1$}), we get
$$\sum_{j=1}^{N}\frac{r_{n+1,j}+r_{n,j}}{z-t_j}
=-2\alpha _n+\sum_{j=1}^{N}R_{n,j}+\sum_{j=1}^{N}\frac{(t_j-\alpha _n)R_{n,j}+\gamma _j}{z-t_j}.$$
Comparing its both sides the coefficients of $(z-t_j)^0$ and $(z-t_j)^{-1}$, we obtain
 \begin{align}
 2\alpha _n&=\sum_{j=1}^{N}R_{n,j},\label{eq25}\\
r_{n+1,j}+r_{n,j}&=(t_j-\alpha _n)R_{n,j}+\gamma _j,\quad j=1, \cdots,N.\label{eq26}
\end{align}

Using (\ref{$S_2$}), we find
\begin{equation}		
	\begin{split}		 1+\sum_{j=1}^{N}\left(r_{n+1,j}-r_{n,j}+\frac{(t_j-\alpha _n)(r_{n+1,j}-r_{n,j})}{z-t_j} \right)\\
		\qquad=2\beta _{n+1}-2\beta _{n}+\sum_{j=1}^{N}\frac{\beta _{n+1}R_{n+1,j}-\beta _{n}R_{n-1,j}}{z-t_j} ,\nonumber
	\end{split}
\end{equation}
from which we obtain the following two equations
\begin{align}
	1+\sum_{j=1}^{N}(r_{n+1,j}-r_{n,j})&= 2(\beta _{n+1}-\beta _n),\label{eq27}\\
	(t_j-\alpha _n)(r_{n+1,j}-r_{n,j})&=\beta _{n+1} R_{n+1,j}-\beta _{n} R_{n-1,j},\quad j=1, \cdots ,N.\label{eq28}
\end{align}
Replacing $n$ by $k$ in (\ref{eq27}) and summing it over from $k=0$ to $n-1$, in view of the fact that $r_{0,j}=\beta_{0}=0$, we are led to
\begin{align}
	n+\sum_{j=1}^{N}r_{n,j}=2\beta _n.\label{eq29} 
\end{align}
Multiplying both sides of (\ref{eq28}) by $R_{n,j}$ and getting rid of the term $(t_j-\alpha_{n})R_{n,j}$ by using (\ref{eq26}), we get
\begin{align}
	r_{n+1,j}^{2}-r_{n,j}^{2}-\gamma _j(r_{n+1}-r_{n,j})=\beta _{n+1}R_{n+1,j}R_{n,j}-\beta _{n}R_{n,j}R_{n-1,j}, \nonumber
\end{align}
for $j=1,\cdots,N$.
Noting that both sides of the above equation are first differences in $n$, we replace $n$ by $k$ in this equation and sum it from $k =0$ to $n-1$. In view of $r_{0,j}=\beta_0=0$, we obtain
\begin{align}
	r_{n,j}^{2}-\gamma _jr_{n,j}=\beta _{n}R_{n,j}R_{n-1,j}, \quad j=1,\cdots,N.\label{eq31}
\end{align}
\begin{remark}
We observe that equations \eqref{eq25} and \eqref{eq29} are exactly the same as \eqref{Ral} and \eqref{rbt}.

	When $N=1$, \eqref{eq25} and \eqref{eq29} are reduced to
	$$R_{n,1}=2\alpha_{n},$$
	$$r_{n,1}=2\beta_{n}-n.$$
Inserting them into \eqref{eq26} and \eqref{eq28}, we get the coupled difference equations for the recurrence  coefficients
    $$\beta _{n+1}+\beta _n=n+\frac{1}{2} +\frac{\gamma _1}{2} +\alpha _n(t_1-\alpha _n),$$
	$$(t_1-\alpha _n)(\beta _{n+1}-\beta_n-\frac{1}{2})=\beta _{n+1}\alpha _{n+1}-\beta _{n}\alpha _{n-1}.$$
These coincide with $(4.5)$ and $(4.6)$ of  \cite{02} respectively.
\end{remark}

Now we summarize the expressions  (\ref{eq25}) and (\ref{eq29}) in the following lemma.
\begin{lemma} \label{3.2}
The recurrence coefficients are expressed in terms of the auxiliary quantities by
\begin{align}
	 \alpha_n&=\frac{1}{2}\sum_{j=1}^{N}R_{n,j},\label{eq32}\\
	\beta _n&=\frac{1}{2} n +\frac{1}{2} \sum_{j=1}^{N}r_{n,j}.\label{eq33}
\end{align}
\end{lemma}

Using the above expressions and the difference equations (\ref{eq26}) and (\ref{eq31}), we establish for the auxiliary quantities a system of difference equations which can be iterated in $n$.
\begin{proposition}
 $\{R_{n,j}, r_{n,j}, j=1,\cdots,N\}$ satisfy the following system of difference equations
\begin{align}
	 r_{n+1,j}=\biggl(t_j-\frac{1}{2}\sum_{j=1}^{N}R_{n,j}\biggr)R_{n,j}+\gamma _j-r_{n,j},\label{eq34} 
\end{align}
\begin{align}
	R_{n,1}=\frac{2r_{n,1}(r_{n,1}-\gamma _1)}{\biggl(n+\sum\limits _{j=1}^{N} r_{n,j}\biggr)R_{n-1,1}}, \label{eq36} 
\end{align}
\begin{align}
	R_{n,j}=\frac{r_{n,j}(r_{n,j}-\gamma _j)}{r_{n,1}(r_{n,1}-\gamma _1)} \cdot \frac{R_{n,1}R_{n-1,1}}{R_{n-1,j}}, \qquad j=2,\cdots ,N,\label{eq35} 
\end{align}
which can be iterated in $n$ with the initial conditions
$$R_{0,j}=\frac{\gamma _j}{h_n}\int_{-\infty }^{+\infty } \frac{{\rm e}^{-x^2} }{x-t_j}\cdot\prod\limits_{k=1}^{N} \left | x-t_k \right | ^{\gamma _k}dx, \qquad r_{0,j}=0,$$
for $j=1,\cdots,N$.
\end{proposition}

\begin{proof}
Substituting (\ref{eq32}) into (\ref{eq26}), we get (\ref{eq34}).
Setting $j=1$ in (\ref{eq31}), we have
\begin{align}
	r_{n,1}^{2}-\gamma _1r_{n,1}=\beta _{n}R_{n,1}R_{n-1,1}.\label{eq05}
\end{align}
Plugging (\ref{eq33}) into the above equation, we come to (\ref{eq36}).
Dividing (\ref{eq31}) for $j=2,\cdots,N$ by (\ref{eq05}), we obtain (\ref{eq35}).
\end{proof}

To close this section, we plug the Taylor expansions \eqref{AnBn-N} for $A_n(z)$ and $B_n(z)$ as $z\rightarrow+\infty$ into
(\ref{$S_{2}'$}) to express $p(n,\vec{t}\,)$, the coefficient of $x^{n-1}$ in $P_n(x;\vec{t}\,)$, in terms of $\{R_{n,j}, r_{n,j}\}$. We will see in Section \ref{A} that this expression plays a vital role in the derivation of the PDE satisfied by the logarithmic derivative of the Hankel determinant \eqref{eq1}.

\begin{lemma}
	$p(n,\vec{t}\,)$ is represented in terms of the auxiliary quantities by
\begin{align}
	 p(n,\vec{t}\,)=\sum_{j=1}^{N}t_jr_{n,j}-\frac{1}{2} \biggl(n+\sum_{j=1}^{N}r_{n,j}\biggr)\sum_{j=1}^{N}R_{n,j}-\sum_{j=1}^{N}\frac{r_{n,j}^2-\gamma_jr_{n,j}}{R_{n,j}}.\label{eq038}
\end{align}
\end{lemma}
\begin{proof}
Plugging \eqref{AnBn-N} into (\ref{$S_{2}'$}), by comparing the coefficients of  $1/z$ on both sides,
 we get
\begin{align}
\sum\limits _{j=1}^{N}t_jr_{n,j}+\sum_{j=0}^{n-1}\alpha _j=2\beta _n(\alpha _n+\alpha _{n-1}), \nonumber
\end{align}
which combined with (\ref{eq034}) gives us
$$p(n,\vec{t}\,)=\sum_{j=1}^{N}t_jr_{n,j}-2\beta_n\alpha_n-2\beta_n\alpha_{n-1}.$$
Replacing $\alpha_{n}$ and $\alpha_{n-1}$ in the above equation by using (\ref{eq32}), in view of (\ref{eq33}), we find
\begin{align}
	 p(n,\vec{t}\,)=\sum_{j=1}^{N}t_jr_{n,j}-\frac{1}{2}\biggl(n+\sum_{j=1}^{N}r_{n,j}\biggr)\sum_{j=1}^{N}R_{n,j}-\sum_{j=1}^{N}\beta _nR_{n-1,j}. \nonumber
\end{align}
Using (\ref{eq31}) to eliminate $\beta_{n}R_{n-1,j}$,  we come to (\ref{eq038}).
\end{proof}

\section{Toda equations, Riccati equations and generalized Painlev\'{e}  IV equation}

We proceed to develop differential relations by differentiating the orthogonality relation (\ref{eq3}) with $m=n$ and $m=n-1$.
By using these relations and the results presented in the preceding section,
we derive Toda equations for the recurrence coefficients and Riccati equations satisfied by the auxiliary quantities.
\begin{lemma}
The relationships between the derivatives of  $\ln{h_n(\vec{t}\,)}$, $p(n,\vec{t}\,)$ and the auxiliary quantities $\{R_{n,j}, r_{n,j} \}$ are given as follows
\begin{align}
		\frac{\partial }{\partial {t_j}} \ln{h_n}&=-R_{n,j}, \label{eq37}\\
	\frac{\partial}{\partial {t_j}} p(n,\vec{t}\, )&=r_{n,j}, \label{eq38}
\end{align}
for $j=1,\cdots ,N.$ Hence, according to \eqref{eq7} and \eqref{eq6}, we find
\begin{align}
     \frac{\partial }{\partial {t_j}} \ln{\beta _n}&=R_{n-1,j}-R_{n,j}, \label{eq39} \\
	\frac{\partial}{\partial {t_j}} \alpha _n&=r_{n,j}-r_{n+1,j}, \label{eq40}
\end{align}
for $j=1,\cdots ,N.$
\end{lemma}

\begin{proof}
Taking the derivative with respect to $t_j$ in the following equation
$$h_n(\vec{t}\, )=\int_{-\infty }^{+\infty } P_{n}^{2}(y;\vec{t}\,)w(y;\vec{t}\,)dy,$$
where $w(y;\vec{t}\,)$ is given by (\ref{eq2}), we get
\begin{equation}
\begin{aligned}
	\frac{\partial }{\partial {t_j}} h_n(\vec{t}\,)=&\int_{-\infty }^{+\infty } 2P_n(y;\vec{t}\,)\cdot\frac{\partial }{\partial {t_j}}P_n(y;\vec{t}\,)\cdot w(y;\vec{t}\, ) dy\\
&+\int_{-\infty }^{+\infty }P_{n}^{2}(y;\vec{t}\,){\rm e}^{-y^2}\biggl(\frac{\partial }{\partial {t_j}} \prod_{k=1}^{N} \left | y-t_k \right | ^{\gamma _k}\biggr)dy.\label{eq06}
\end{aligned}
\end{equation}
Applying $\frac{\partial }{\partial {t_j}} $ to
$$P_n(y;\vec{t}\, )=y ^n+p(n,\vec{t}\, )y^{n-1}+\cdots ,$$
we know that the degree of $\frac{\partial }{\partial {t_j}}P_n(y;\vec{t}\,)$ is $n-1$. Hence
the first integral on the right hand side of (\ref{eq06}) is zero and consequently
\begin{equation}		
	\begin{split}
		\frac{\partial }{\partial {t_j}} h_n(\vec{t}\,)&=\int_{-\infty }^{+\infty }P_{n}^{2}(y;\vec{t}\,){\rm e}^{-y^2}\biggl(\frac{\partial }{\partial {t_j}} \prod_{k=1}^{N} \left | y-t_k \right | ^{\gamma _k}\biggr)dy\\
		&=-\gamma _j\int_{-\infty }^{+\infty }\frac{P_{n}^{2}(y;\vec{t}\,)}{y-t_j} w(y;\vec{t}\,) dy\nonumber\\
&=-h_nR_{n,j},
	\end{split}		
\end{equation}
where the second equality is obtained by using an argument similar to the derivation of (\ref{eq15}) and the third one is due to the definition of $R_{n,j}$ given by (\ref{defRn-1}). This completes the proof of (\ref{eq37}).

To continue,  we differentiate the orthogonality relation
$$0=\int_{-\infty }^{+\infty } P_n(y;\vec{t}\,)P_{n-1}(y;\vec{t}\,)w(y;\vec{t}\, ) dy$$
over $t_j$, and find
\begin{equation}		
	\begin{split}
		0=&\int_{-\infty }^{+\infty } P_n(y;\vec{t}\,)\cdot\frac{\partial }{\partial {t_j}}P_{n-1}(y;\vec{t}\,)\cdot w(y;\vec{t}\, ) dy\\
&+\int_{-\infty }^{+\infty } \frac{\partial }{\partial {t_j}}P_n(y;\vec{t}\,)\cdot P_{n-1}(y;\vec{t}\,)w(y;\vec{t}\, ) dy\\&+\int_{-\infty }^{+\infty }P_{n}(y;\vec{t}\,)P_{n-1}(y;\vec{t}\,){\rm e}^{-y^2}\biggl(\frac{\partial }{\partial {t_j}} \prod_{k=1}^{N} \left | y-t_k \right | ^{\gamma _k}\biggr)dy. \label{eq07}
	\end{split}		
\end{equation}
Noting that $\frac{\partial }{\partial {t_j}}P_{n-1}(y;\vec{t}\, )$ is of degree $n-2$, we see from the orthogonal relation (\ref{eq3}) that the first term on the right hand side of the above equation is zero.
Observing that
$$\frac{\partial }{\partial {t_j}}P_n(y;\vec{t}\, )=\frac{\partial }{\partial {t_j}}(y ^n+p(n,\vec{t}\, )y^{n-1}+\cdots)=\frac{\partial }{\partial {t_j}}p(n,\vec{t}\, )y^{n-1}+\cdots,$$
in view of (\ref{eq3}), we find that the second integral in \eqref{eq07} is equal to $h_{n-1}\frac{\partial}{\partial {t_j}} p(n,\vec{t}\, )$. Via an argument similar to the derivation of (\ref{eq15}), in light of the definition of $r_{n,j}$ given by (\ref{defrn-1}), we find that the third integral in \eqref{eq07} is $h_{n-1}r_{n,j}$. Hence, (\ref{eq07}) becomes
\begin{equation*}		
	\begin{split}
		0&=h_{n-1}\frac{\partial}{\partial {t_j}} p(n,\vec{t}\, )-h_{n-1}r_{n,j},
	\end{split}	
\end{equation*}
which gives us (\ref{eq38}).
\end{proof}

According to (\ref{eq39})-(\ref{eq40}) and (\ref{eq32})-(\ref{eq33}), we arrive at the following Toda equations for the recurrence coefficients.
\begin{proposition}
The recurrence coefficients satisfy the following Toda equations
\begin{align}
	&\delta \ln \beta_n=2(\alpha_{n-1}-\alpha_n),\label{eq08}\\
	&\delta \alpha _n=1+2(\beta _{n}-\beta _{n+1}), \label{eq09}
\end{align}
where $\delta =\sum\limits_{j=1}^{N} \frac{\partial }{\partial {t_j}}.$
\end{proposition}
\begin{proof}
Summing (\ref{eq39}) 
over $j$ from $1$ to $N$,  we have
\begin{equation}		
	\begin{split}
		\sum_{j=1}^{N}\frac{\partial }{\partial {t_j}} \ln\beta _n
		 &=\sum_{j=1}^{N}R_{n-1,j}-\sum_{j=1}^{N}R_{n,j}\\
		&=2(\alpha_{n-1}-\alpha_n), \nonumber
	\end{split}	
\end{equation}
which gives us (\ref{eq08}). Here note that the second equality above is due to (\ref{eq32}).

Similarly, summing (\ref{eq40})
over $j$ from $1$ to $N$, in view of \eqref{eq33}, we get
\begin{equation}		
	\begin{split}
		\sum_{j=1}^{N}\frac{\partial }{\partial {t_j}}\alpha _n
		 &=\sum_{j=1}^{N}r_{n,j}-\sum_{j=1}^{N}r_{n+1,j}\\
		&=1+2(\beta _{n}-\beta _{n+1}),\nonumber
	\end{split}	
\end{equation}
which gives us (\ref{eq09}).
\end{proof}

Now we proceed to derive the Riccati equations satisfied by the auxiliary quantities $\left\lbrace R_{n,j}, r_{n,j}\right\rbrace $. To do this, we combine the differential relations for the recurrence coefficients given by (\ref{eq39}) and (\ref{eq40}) with the difference identities and expressions found in the preceding section.
\begin{theorem}
	The auxiliary quantities $\left\lbrace R_{n,j}, r_{n,j},j=1,\cdots ,N\right\rbrace $ satisfy the following Riccati equations
\begin{align}
	\delta R_{n,j}&=4r_{n,j}-\biggl(2t_j-\sum_{k=1}^{N}R_{n,k}\biggr)R_{n,j}-2\gamma _j, \label{eq41} \\
   \delta r_{n,j}&=\frac{2r_{n,j}(r_{n,j}-\gamma _j)}{R_{n,j}}-\biggl(n+\sum_{k=1}^{N}r_{n,k}\biggr)R_{n,j}, \label{eq42}
\end{align}
for $j=1,\cdots,N$, where $\delta =\sum\limits_{k=1}^{N} \frac{\partial }{\partial {t_k}}.$
\end{theorem}

\begin{proof}
	From (\ref{eq37}), it follows that
	$$\frac{\partial^2}{\partial {t_j}\partial{t_k}} \ln{h_n} =-\frac{\partial}{\partial {t_k}} R_{n,j},$$
	$$\frac{\partial^2}{\partial {t_k}\partial{t_j}} \ln{h_n} =-\frac{\partial}{\partial {t_j}} R_{n,k},$$
for $j,k=1,\cdots,N$. Since
$\frac{\partial^2}{\partial {t_j}\partial{t_k}} \ln{h_n} =\frac{\partial^2}{\partial {t_k}\partial{t_j}} \ln{h_n}$,
 we find
 \begin{align}
 	\frac{\partial}{\partial {t_j}} R_{n,k}=\frac{\partial}{\partial {t_k}} R_{n,j}, \qquad j,k=1,\cdots,N.\label{010}
 \end{align}
   Similarly, using (\ref{eq38}) and the fact that
   $\frac{\partial^2}{\partial {t_j}\partial{t_k}} p(n,\vec{t}\, ) =\frac{\partial^2}{\partial {t_k}\partial{t_j}} p(n,\vec{t}\, )$, we obtain
\begin{align}
	\frac{\partial}{\partial {t_j}} r_{n,k}=\frac{\partial}{\partial {t_k}} r_{n,j}, \qquad j,k=1,\cdots,N.\label{eq011}
\end{align}

Now we go ahead with the derivation of the Riccati equations satisfied by $R_{n,j}$ and $r_{n,j}$.
Inserting (\ref{eq32}) into (\ref{eq40}) and using (\ref{eq26}) to eliminate $r_{n+1,j}$ in the resulting equation, we get
 \begin{align}
\frac{1}{2}\sum_{k=1}^{N}\frac{\partial}{\partial {t_j}}R_{n,k}=2r_{n,j}-(t_j-\alpha _n)R_{n,j}-\gamma _j. \nonumber
 \end{align}
Replacing in the above equality $\frac{\partial}{\partial {t_j}} R_{n,k}$ by $\frac{\partial}{\partial {t_k}} R_{n,j}$ which is due to (\ref{010}), in view of (\ref{eq32}), we obtain (\ref{eq41}).

Using (\ref{eq39}) to eliminate $R_{n-1,j}$ in
 (\ref{eq31}), we have
 \begin{align}
   r_{n,j}^{2}-\gamma _jr_{n,j}=R_{n,j}\biggl(\frac{\partial }{\partial {t_j}}\beta _n+\beta _nR_{n,j} \biggr). \label{eq47}
\end{align}
Taking the derivative on both sides of (\ref{eq33}) with respect to $t_j$, in view of \eqref{eq011}, we get
$$	\frac{\partial }{\partial {t_j}} \beta _n=\frac{1}{2}\frac{\partial }{\partial{t_j} }\sum_{k=1}^{N}r_{n,k}=\frac{1}{2}\delta r_{n,j}. $$
Substituting it and (\ref{eq33}) into (\ref{eq47}) leads us to (\ref{eq42}).
\end{proof}

Solving $r_{n,j}$ from (\ref{eq41}) and substituting it into (\ref{eq42}), noting that $\delta (t_j) =1$,
after simplification, we arrive at the following PDEs satisfied by $\{R_{n,j}\}$.
\begin{theorem}
The auxiliary quantities $\{R_{n,j}(\vec{t}\,),j=1,\cdots,N\}$ satisfy the following second order non-linear PDEs
\begin{equation}
	\begin{aligned}
	\frac{1}{2}\delta ^{2} R_{n,j}=&\frac{(\delta R_{n,j})^2}{4R_{n,j}} -\left[\sum_{k=1}^{N}\biggl(\biggl(t_k-\frac{1}{2} \sum_{k=1}^{N}R_{n,k}\biggr)R_{n,k}+\gamma _k\biggr)\right]R_{n,j}\\	 &+\biggl(t_j-\frac{1}{2}\sum_{k=1}^{N}R_{n,k}\biggr)^2R_{n,j}-(2n+1)R_{n,j}-\frac{\gamma _{j}^{2}}{R_{n,j}},\label{eq48}
	\end{aligned}
\end{equation}	
for $j=1,\cdots,N$, where $\delta =\sum\limits_{k=1}^{N} \frac{\partial }{\partial {t_k}}$ and $\delta^2=\sum\limits_{j=1}^{N}\sum\limits_{k=1}^{N}\frac{\partial^2}{\partial t_j\partial t_k}$.
\end{theorem}

\begin{remark}
	When $N=1$,	the system \eqref{eq48} is reduced to a second order ODE for $R_{n,1}(t_1):$
\begin{equation}\label{Rode-1}
	\begin{aligned}
	R_{n,1}''=&\frac{(R_{n,1}')^2}{2R_{n,1}} -\left[\left(2t_1-R_{n,1}\right)R_{n,1}+2\gamma _1\right]R_{n,1}\\	 &+2\left(t_1-\frac{1}{2}R_{n,1}\right)^2R_{n,1}-2(2n+1)R_{n,1}-\frac{2\gamma _{1}^{2}}{R_{n,1}}.
	\end{aligned}
\end{equation}	
Since it follows from \eqref{eq32} with $N=1$ that $R_{n,1}(t_1)=2\alpha_n(t_1)$, we readily get the ODE for $\alpha_n(t_1)$ from \eqref{Rode-1}, which coincides with $(4.17)$ of \cite{02}. In addition, by setting $R_{n,1}(t_1)=:R_n(t)$ with $t_1=:-t$ in \eqref{Rode-1}, we find
\begin{align}
 R_{n}''(t)=\frac{(R_{n}'(t))^2}{2R_{n}(t)}+\frac{3}{2}R_{n}^3(t)+4tR_{n}^2(t)+2(t^2-2n-1-\gamma_1)R_{n}(t)-\frac{2\gamma_1^2 }{R_{n}(t)}, \nonumber
\end{align}
which agrees with $(5.1)$ of \cite{02} and was identified to be a  Painlev\'{e} IV equation \cite{18} with $\alpha=2n+1+\gamma_1$, $\beta=-2\gamma_1^2$.
\end{remark}

\section{Generalized $\sigma$-form of Painlev\'{e}  IV equation} \label{A}

In this section, we focus on the derivation of the PDE satisfied by the logarithmic derivative of the Hankel determinant $D_n(\vec{t} \,)$ given by (\ref{eq1}).
Define
\[\sigma_n(\vec{t}\,):=\delta \ln{D_n(\vec{t}\,)},\]
where $\delta =\sum\limits_{k=1}^{N} \frac{\partial }{\partial {t_k}}$.
A combination of (\ref{eq033}) and (\ref{eq37}) gives us
\begin{align}
	 \sigma_n(\vec{t}\,)=&-\sum_{j=0}^{n-1}\sum_{k=1}^{N}R_{j,k} \nonumber\\
	=&-2\sum_{j=0}^{n-1}\alpha_j \nonumber\\
	=&2p(n,\vec{t}\,),\label{eq036}
\end{align}
where the second equality is due to (\ref{eq32}) and the third one results from (\ref{eq034}).

Using these relations and identities in the previous sections, we represent $\sigma_n$ and the auxiliary quantities $\{R_{n,j}, r_{n,j}\}$ by each other.
\begin{theorem}
	$\sigma_n$ is expressed in terms of the auxiliary quantities $\{R_{n,j}, r_{n,j}, j=1,\cdots,N\}$ by
\begin{align}
	 \sigma_n=2\sum_{j=1}^{N}t_jr_{n,j}-\biggl(n+\sum_{j=1}^{N}r_{n,j}\biggr)\sum_{j=1}^{N}R_{n,j}-2\sum_{j=1}^{N}\frac{r_{n,j}^2-\gamma_jr_{n,j}}{R_{n,j}}.\label{eq025}
\end{align}
The auxiliary quantities $\{R_{n,j}, r_{n,j}\}$ are expressed in terms of $\sigma_n$ and its derivatives by
\begin{align}
	r_{n,j}=\frac{1}{2}\cdot\frac{\partial \sigma_n}{\partial{t_j}},\label{eq026}
\end{align}
\begin{align}
	R_{n,j}=\frac{1}{2(2n+\delta \sigma_n)} \biggl [ -\biggl(\frac{\partial^2\sigma_n}{\partial{t_j^2}}+\sum_{\substack{k=1 \\ k\neq j}}^N \frac{\partial^2\sigma_n}{\partial{t_k}\partial{t_j}} \biggr ) +{\rm sgn}(R_{n,j}+R_{n-1,j}) \sqrt{\Delta_j(\vec{t}\, )} \biggr ],\label{eq027}
\end{align}
for $j=1,\cdots,N$, where $\delta =\sum\limits_{k=1}^{N} \frac{\partial }{\partial {t_k}}$ and ${\rm sgn}(R_{n,j}+R_{n-1,j})$ is the sign function of $R_{n,j}+R_{n-1,j}$ which is $-1$ for $R_{n,j}+R_{n-1,j}<0$, $1$ for $R_{n,j}+R_{n-1,j}>0$ and $0$ for $R_{n,j}+R_{n-1,j}=0$.
Here $\Delta_j(\vec{t}\, )$ is defined by
\begin{align}
	\Delta_j(\vec{t}\, ):=   \left(\frac{\partial^2\sigma_n}{\partial{t_j^2}}+\sum_{\substack{k=1 \\ k\neq j}}^N \frac{\partial^2\sigma_n}{\partial{t_k}\partial{t_j}} \right )^2+4(2n+\delta \sigma _n)\frac{\partial\sigma_n}{\partial{t_j}}\left ( \frac{\partial\sigma_n}{\partial{t_j}}-2\gamma _j \right ),\label{eq028}
\end{align}
for $j=1,\cdots,N$.
\end{theorem}
\begin{proof}
Expression (\ref{eq025}) is an immediate result of (\ref{eq038}) and (\ref{eq036}).
Differentiating both sides of  (\ref{eq036}) with respect to $t_j$, in view of (\ref{eq38}), we have
$$\frac{\partial \sigma_n}{\partial{t_j}} =2r_{n,j},$$
which gives us (\ref{eq026}).

Now we proceed with the derivation of (\ref{eq027}).
By rewriting  (\ref{eq42}), we come to the following quadratic algebraic equation in $R_{n,j}$
\begin{align}
	 \biggl(n+\sum_{k=1}^{N}r_{n,k}\biggr)R_{n,j}^2+\delta r_{n,j}\cdot R_{n,j}-2r_{n,j}(r_{n,j}-\gamma_j)=0. \label{eq050}
\end{align}
Now we discuss whether this equation has real roots or not by looking at its discriminant
\begin{align*}
	\tilde{\Delta}_j(\vec{t}\,):=(\delta r_{n,j})^2+8\biggl(n+\sum_{k=1}^{N}r_{n,k}\biggr)r_{n,j}(r_{n,j}-\gamma_j).
\end{align*}
Plugging the Riccati equation (\ref{eq42}) into the above expression, we get
\begin{align*}
	 \tilde{\Delta}_j(\vec{t}\,)=\biggl[\frac{2r_{n,j}(r_{n,j}-\gamma_j)}{R_{n,j}}+\biggl(n+\sum_{k=1}^{N}r_{n,k}\biggr) R_{n,j}\biggr]^2\ge 0.
\end{align*}
This implies that (\ref{eq050}) has real roots given by
\begin{align}
	 R_{n,j}=\frac{ -\delta r_{n,j}\pm \sqrt{\tilde{\Delta}_j}}{2\biggl(n+\sum\limits_{k=1}^{N}r_{n,k}\biggr)}. \label{eq031}
\end{align}

Next we determine the sign before the square root in (\ref{eq031}). Taking the derivative on both sides of (\ref{eq33}) with respect to $t_j$,
in view of (\ref{eq011}), we find
\begin{align}
	\frac{\partial \beta _n}{\partial {t_j}}=\frac{1}{2}\delta r_{n,j}.\nonumber
\end{align}
Using it and (\ref{eq33}) to get rid of $\delta r_{n,j}$ and $n+\sum\limits_{k=1}^{N}r_{n,k}$ in (\ref{eq031}), we are led to
\begin{align}
	R_{n,j}=\frac{1}{4\beta_n}\biggl (-2\frac{\partial \beta _n }{\partial {t_j}}\pm \sqrt{\tilde{\Delta}_j } \biggr ).\nonumber
\end{align}
Eliminating the term $\frac{\partial{\beta _n} }{\partial {t_j}}$ in the above equation by using (\ref{eq39}), we obtain
$$R_{n,j}+R_{n-1,j}=\pm \frac{1}{2\beta_n}\sqrt{\tilde{\Delta}_j},$$
which indicates that the sign before the above square root  is ${\rm sgn}(R_{n,j}+R_{n-1,j})$. Consequently, (\ref{eq031}) becomes
\begin{align}
	 R_{n,j}=\frac{ -\delta r_{n,j}+{\rm sgn}(R_{n,j}+R_{n-1,j}) \sqrt{\tilde{\Delta}_j} \,}{2\left(n+\sum\limits_{k=1}^{N}r_{n,k}\right)}. \label{eq053}
\end{align}
Inserting (\ref{eq026}) into the above equality and writing $\Delta_j(\vec{t}\,):=4\tilde{\Delta}_j(\vec{t}\,)$, we arrive at (\ref{eq027}).
\end{proof}

Substituting (\ref{eq026}) and (\ref{eq027}) back into (\ref{eq025}), after simplification, we obtain the PDE satisfied by $\sigma_n(\vec{t}\,)$.
\begin{theorem}
$\sigma_n(\vec{t}\,)$ satisfies the following second order PDE
\begin{align}
	\sigma_n(\vec{t}\,)= \sum_{j=1}^{N}t_j\frac{\partial\sigma_n }{\partial{t_j}}-\frac{1}{2}\sum_{j=1}^{N}{\rm sgn}(R_{n,j}+R_{n-1,j}) \sqrt{\Delta_j(\vec{t}\, )}, \label{eq029}
\end{align}
 where $\Delta_j(\vec{t}\, )$ is given by \eqref{eq028}.
\end{theorem}
\begin{proof}
Before plugging (\ref{eq026}) and (\ref{eq027}) into (\ref{eq025}), to simplify the calculations, we first rewrite the quantity $\sum\limits_{j=1}^{N}\frac{r_{n,j}^2-\gamma_jr_{n,j}}{R_{n,j}}$ that appears in (\ref{eq025}). According to (\ref{eq31}) and (\ref{eq39}), we find
\begin{align}
	 \sum_{j=1}^{N}\frac{r_{n,j}^2-\gamma_jr_{n,j}}{R_{n,j}}=&\sum_{j=1}^{N}\beta _nR_{n-1,j}\nonumber \\
	=&\delta \beta _n+\beta _n\sum_{j=1}^{N}R_{n,j}. \label{eq052}
\end{align}
Inserting (\ref{eq052}) back into (\ref{eq025}), in view of (\ref{eq33}), we get
\begin{align}
	 \sigma_n&=2\sum_{j=1}^{N}t_jr_{n,j}-2\biggl(n+\sum_{j=1}^{N}r_{n,j}\biggr) \sum_{j=1}^{N}R_{n,j}-2\delta \beta_n \nonumber\\
	 &=2\sum_{j=1}^{N}t_jr_{n,j}+\sum_{j=1}^{N}\delta r_{n,j}-\sum_{j=1}^{N}{\rm sgn}(R_{n,j}+R_{n-1,j}) \sqrt{\tilde{\Delta}_j}-2\delta \beta _n, \label{eq054}
\end{align}
where the second equation is obtained by substituting (\ref{eq053}) into the first one. Here note that $\tilde{\Delta}_j=\Delta _j/4$ with $\Delta _j$ defined by (\ref{eq028}). Plugging (\ref{eq026}) into (\ref{eq054}), noting that $\sum\limits_{j=1}^{N}\delta r_{n,j}=2\delta \beta _n$ which results from (\ref{eq33}), we finally come to (\ref{eq029}).
\end{proof}
\begin{remark}
When $N=1$, \eqref{eq029} becomes after clearing the square root
$$(\sigma_n'')^2=4(t_1\sigma_n'-\sigma_n)^2-4\sigma_n'(\sigma_n'-2\gamma_1)(\sigma_n'+2n),$$
which agrees with $(2.39)$ of \cite{03}. As was pointed out therein, this equation is the $\sigma$-form of a Painlev\'{e}  IV equation \cite{19} with ${\nu}_0=0$, $\nu_1=-2\gamma_1$ and $\nu_2=2n$.
\end{remark}

\section*{Acknowledgments}
This work was supported by National Natural Science Foundation of China under grant number 12101343 and by Shandong Provincial Natural Science Foundation with project number ZR2021QA061.

\end{document}